\documentclass[aps,pre,twocolumn,showpacs,eqsecnum]{revtex4}

\usepackage{graphicx}
\usepackage{bm}
\usepackage{amsmath}
\usepackage{dcolumn}

\begin{document}

\bibliographystyle{apsrev}

\title {Low-frequency plasma conductivity in the average-atom approximation}

\author{M.Yu.Kuchiev} \email[Email:]{kuchiev@newt.phys.unsw.edu.au}
 \affiliation {Department of Theoretical Physics, School of Physics, University of New South Wales,
 Sydney, 2052, Australia}

\author{W.R.Johnson} \email[Email:]{johnson@nd.edu}
\affiliation {Department of Physics, 225 Nieuwland Science Hall,
University of Notre Dame, Notre Dame IN 46556}

        \date{\today}
\begin{abstract}
     Low-frequency properties of a plasma are examined within the
     average-atom approximation, which presumes that scattering of a
     conducting electron on each atom takes place
     independently of other atoms. The relaxation time $\tau$
     distinguishes a high-frequency region $\omega \tau >
     1$, where the single-atom approximation is applicable explicitly,
     from extreme low frequencies $\omega \tau < 1 $, where,
     naively, the single-atom approximation is invalid.  A proposed
     generalization of the formalism, which takes into account
     many-atom collisions, is found to be accurate in all
     frequency regions, from $\omega =0$ to $\omega\tau>1$,
     reproducing the Ziman formula in the static limit, results
     based on the Kubo-Greenwood formula for high frequencies, and
     satisfying the conductivity sum-rule precisely.  The correspondence between physical
     processes leading to the conventional Ohm's law and the infrared
     properties of QED is discussed. The suggested average-atom approach to frequency-dependent
     conductivity is illustrated by numerical calculations for the an aluminum plasma
     in the temperature range 2--10 eV.
\end{abstract}


\maketitle

\section{Introduction}
   \label{intro}

  Various theoretical approaches are available to investigate the frequency-dependent conductivity
   of plasmas,
  ranging from methods based on a many-body expansion of the grand canonical
  partition function \cite{IRW:87,RWI:88,IR:96} to methods based on molecular dynamics
  simulations \cite{DKC:02,RCR:03,CRL:05,D:05,FBR:06,FBR:06b}.  In the present
  paper, we re-examine an average-atom approach \cite{johnson_guet_bertsch_04}
  that has been used recently to investigate anomalous dispersion in C, Al, Ag, and other
  plasmas in the soft x-ray region (14-47 nm) of the spectrum
  \cite{NJ:05,NJI:06,JJM:06,NJC:07,JJM:07}.
  The utility of the average atom method rests on its simplicity and wide range of applicability.

    At low
   frequencies,  electron-ion scattering contributions dominate
   the conductivity $\sigma(\omega)$, while at higher frequencies (e.g. in the x-ray region mentioned above)
   photoionization and bound-bound transitions provide the most important
   contributions. Effects of multiple scattering
   were omitted in evaluating free-free contributions to
   $\sigma(\omega)$
   in Ref.~\cite{johnson_guet_bertsch_04}, leading to a (spurious) second-order pole
   at $\omega=0$ that was regularized in an {\it ad hoc} way.
   In the paragraphs below, we discuss the origin of this
   pole in more detail and a give a modified formula for the free-free contribution
   to $\sigma(\omega)$ that accounts for multiple scattering, is regular at
    $\omega=0$, and  rigorously satisfies the conductivity sum rule.

   The present discussion concerns the plasma conductivity $\sigma(\omega)$ at
   low frequencies, i.e.\ presuming that the frequency $\omega$ is
   lower than both the plasma frequency and typical frequencies of
   atomic excitations
   \begin{equation}\label{freq}
     \omega \ll (4 \pi n_e e^2/m)^{1/2}, ~me^4/\hbar^3~.
   \end{equation}
   The only parameter that drives the conductivity in this region is
   the relaxation time $\tau$, which establishes a boundary between
   relatively high-frequencies $\omega \tau > 1$, which will be called
   the {\it high-frequencies} for short, and extreme low-frequencies,
   where $\omega \tau < 1$ including the static limit $\omega=0$;
   these frequencies will be called the {\it ultra-low frequencies}.

   Physical processes, which govern the conductivity in these two
   regions, differ qualitatively, as discussed in detail
   below.  Alongside this physical difference, there exists also a
   distinction in theoretical methods. One line of research is based
   on the Ziman formula, which is applicable in the static limit,
   leading to the conventional static Ohm's law, see
   \citet{ziman_1964} ch.7,  \citet{mahan_2000} ch.8. An alternative
   approach is based on the Kubo-Greenwood formula
   \cite{kubo_56,kubo_57,greenwood_58,harrison_1970,ashcroft-mermin_1976},
   which usually gives a reliable description of conductivities at high-frequencies.

   Generally speaking, the Kubo-Greenwood formalism should lead to
   accurate results for arbitrary frequencies, provided though
   that all {\it important} scattering processes are taken into
   account.  However, typically, within some given theoretical scheme,
   it is feasible to account only for some particular class of
   scattering events.  This restriction may substantially reduce an
   area of applicability for the Kubo-Greenwood formalism.  In
   particular, it is usually difficult to extend its validity to the static
   limit.  An example illustrating the latter
   fact is provided by models based on the  average-atom approximation.
   In such models,
   scattering of conducting electrons is assumed to take
   place on each atom, independently of other atoms.  This
   means that many-atom events (multiple scattering), in which several atoms
   produce a coherent contribution are neglected.  The simplicity and
   clear physical nature of the average-atom approximation make it popular.
   Its origins can be traced to the Thomas-Fermi model of plasma devised
   more than a half century ago by Feynman, Metropolis, and Teller
   \cite{feynman-metropolis-teller_1949}. A quantum mechanical version of
   the  average-atom model is given by 
   \citet{blenski-ishikawa_1995} and a
    recent implementation is found in
    Ref.~\cite{johnson_guet_bertsch_04}.

   The present work shows that scattering processes which take place
   at ultra-low frequencies necessarily include several atoms.
   We will call these processes, the {\it many-atom}
   collisions for short. Their importance indicates that for ultra-low
   frequencies the single-atom approximation breaks down. This fact
   explains a difficulty that occurs in the Kubo-Greenwood formalism in
   the static limit.  The breakdown of the single-atom approximation
   manifests itself as a divergence of the conductivity calculated in
   the Kubo-Greenwood formalism in the limit $\omega \rightarrow 0$.
   As mentioned in the introduction,
   the conductivity in this limit exhibits a second-order pole.

   Thus, direct numerical calculations based on an average-atom
   approximation and relying on the Kubo-Greenwood approach are
   applicable for high frequencies only, while for lower frequencies,
   where the many-atom collisions are important, the formalism faces
   a difficulty.  This work resolves this difficulty, proposing a new
   approach that is applicable for frequencies that satisfy
   the inequality (\ref{freq}). In the static limit $\omega=0$, our description
   reproduces the Ziman formula.  For high frequencies $\omega \tau >
   1$, our results agree with the conventional Kubo-Greenwood
   description. In the intermediate region $\omega \tau \simeq 1$, the
   validity of our formalism is supported by the fact that it provides
   the correct result for the conductivity sum-rule.
   One of the important advantages of the
   proposed description is related to its simplicity.  We show that
   all necessary physical quantities can be evaluated using a simple
   single-atom approximation. This means that multiple scattering,
   which is paramount in the static limit, is accounted for effectively
   in the single-atom approximation!

\begin{figure}[thb]
  \centering \includegraphics[ height=2.8cm, keepaspectratio =
  true,angle =0]{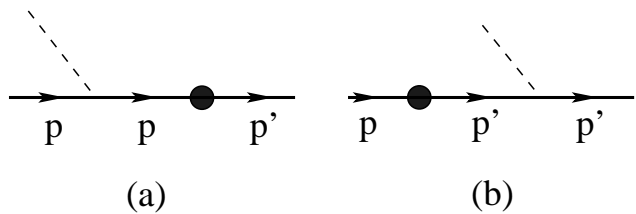}
  \caption{ \label{one}
    Two Feynman diagrams represent the amplitude, which describes
    electron scattering with absorption of a quantum of the
    electromagnetic field, which has frequency $\omega$ and
      small wave vector that leaves the electron momentum unchanged.
      The solid lines show the electron
    propagation, the dashed line - the quantum of the electromagnetic
    field, the filled dot - the elastic scattering process. In
    diagrams (a) and (b) the lines that represent the
    electromagnetic quantum are inserted into the outer legs, which
    makes these diagrams infrared singular, $\propto 1/\omega$ when
    $\omega \rightarrow 0$.  Other possible diagrams have no such
    singularity.}
    \end{figure}

   There is an important relation between the divergence in the
   conductivity at ultra-low frequencies and the infrared problem of
   QED. To make this point more transparent, let us keep in mind that
   the conductivity describes absorption and emission of quanta
   of the electromagnetic field, which are possible due to electron
   scattering.  Presuming that the potential, which is responsible for
   scattering, is localized in a vicinity of some atom, one can
   express the amplitude of absorption $f_\mathrm{abs}$ (or
   emission) in terms of the elastic scattering amplitude $f$.
   This relation reveals that the absorption amplitude has a pole
   at $\omega = 0$
   \begin{equation}
     \label{pole}
     f_\mathrm{abs} \propto  \frac{f} {\omega}~,\quad\quad \omega \rightarrow 0~.
   \end{equation}
   This general, well-known, feature of the infrared processes in QED,
   is described by Feynman diagrams with a photon line inserted
   into the outer electron legs as is shown in Fig.~\ref{one}. The
   first-order pole in the absorption amplitude in Eq.(\ref{pole})
   leads to a second-order pole in the conductivity,
   \begin{eqnarray}
     \label{sec}
     \sigma(\omega) \propto \frac{1}{\omega^2}~.
   \end{eqnarray}
   Developing this argument, we will show below that many-atom
   collisions prevent the divergence of the scattering amplitude in
   Eq.(\ref{pole}) at $\omega \rightarrow 0$. This happens because
   many-atom collisions lead to a finite relaxation time $\tau$, which
   measures the interval of time during which the electron travels
   between two subsequent collisions with different atoms.  We show
   that the relaxation time provides an effective cutoff for the
   amplitude in Eq.(\ref{pole}), in which the pole is replaced by a
   finite quantity $|f_\mathrm{abs}|\simeq 1/\omega_\mathrm{min} =
   \tau$. The well defined, finite, scattering amplitude leads to
   a conductivity that is regular at $\omega=0$; Eq.(\ref{sec}) is replaced by the relation
   $\sigma(\omega) \propto 1/\omega^2_\mathrm{min}=\tau^2$.

    Our discussion below presumes that the plasma conductivity is due
    mainly to scattering of conducting electrons by atomic cores.
    There exist other mechanisms contributing to conductivity. One of
    them is related to electron-electron scattering. The main idea
    of this work can be generalized to cover this mechanism (and others)  as well.
    However, in order to keep our presentation
    simple and clear, we will not attempt to formulate the idea in the most
    general case, restricting our discussion to electron-atom
    scattering only. At sufficiently high temperatures, when atomic
    cores are highly ionized, one expects that electron-atom
    scattering gives the dominant contribution to the conductivity
    owing to the fact that scattering by an ion is a coherent process,
    with probability proportional to $Z_\mathrm{ion}^2$, where
    $Z_\mathrm{ion}$ is the ionic charge. By contrast,
    electron-electron scattering is an incoherent process with
    probability proportional to $Z_\mathrm{ion}$. Thus, scattering
    by an ions is expected to dominate electron-electron scattering,
    provided $Z_\mathrm{ion}>1$.

   \section{Single-atom approximation for high frequencies}

   \subsection{Absorption of photons and elastic scattering}
   We need to recall several simple important facts related to
   absorption of low-frequency quanta by electrons.  Let us presume
   that there is a localized potential $U=U(\mathrm{r})$, which causes
   electron scattering.  Let us assume further that there is some
   external low-frequency homogeneous electric field. \cite{endnote1}
   Then scattering can be accompanied by absorption of a quantum of
   the electromagnetic field. The process of absorption is described
   by the matrix element $f_\mathrm{abs}$,
   \begin{equation}
     \label{me}
     f_\mathrm{abs}=\langle \psi_f| \bm{\epsilon} \cdot \bm{p}\,|
     \psi_i\,\rangle~.
   \end{equation}
   Here $\bm{\epsilon}$ and $\bm{p}$ are the
   polarization vector of the electromagnetic quantum and the operator
   of momentum of the electron, $\psi_i$ and $\psi_f$ are the wave
   functions of the electron in the initial and final states. We are
   interested in the low frequency region specified by
   Eq.(\ref{freq}). Our first goal is to simplify the matrix element
   in Eq.(\ref{me}), presenting it as
   \begin{equation}
     \label{si}
     f_\mathrm{abs}=\frac{ \bm{\epsilon}   \cdot
       \bm{p} }{\hbar \omega}
     \langle \psi_{\mathbf{p}'}|\,U|\,\mathbf{p}\rangle -
     \frac{\bm{\epsilon} \cdot \mathbf{p}' }{\hbar \omega}
     \langle \mathbf{p}'|\,U|\,\psi_\mathbf{p}\rangle
   \end{equation}
   Here the first and second terms correspond to the Feynman diagrams,
   in which the line representing the electromagnetic quantum is
   inserted into the left and right legs of the diagram respectively,
   see Fig.~\ref{one}.  The wave-functions $|\mathbf{p}\rangle$ and
   $|\psi_\mathbf{p}\rangle$ in Eq.(\ref{si}) describe the electron
   propagation in the plane wave approximation and with account of the
   potential $U$, respectively.  All processes, in which the line
   representing the electromagnetic quantum is inserted into
   internal parts of the diagram have no poles in the limit $\omega
   \rightarrow 0$, allowing one to neglect them in Eq.(\ref{si}) (see
   the more detailed discussion after Eq.(\ref{qu})).

   Generally speaking, the electron energy in the initial and final
   states of the photo-absorption process are different.
   However, for low frequencies this difference is insignificant.
   Neglecting it, one can presume that the matrix elements in
   Eq.(\ref{si}) are related to elastic events, i.e.\
   $|\mathbf{p}|=|\mathbf{p}'|$.  Remember now that the elastic
   scattering amplitude is defined as
   \begin{equation}
     \label{el}
f=-\frac{m}{2\pi\hbar^2}
\langle \psi_{\mathbf{p}'}|\,U\,|\mathbf{p}\rangle=
-\frac{m}{2\pi\hbar^2}
\langle \mathbf{p}'|\,U|\,\psi_\mathbf{p}\rangle~,
    \end{equation}
    Consequently one finds from Eq.(\ref{si}) that
   \begin{equation}
     \label{qu}
f_\mathrm{abs}=\frac{2\pi \hbar}{m\omega} \,(\bm{\epsilon} \cdot
\mathbf{q})\, f
   \end{equation}
   where $ \bm{q}= \bm{p}'-\bm{p}$
   is the transferred momentum.

   Eq.(\ref{qu}) relates the amplitude of the process with absorption
   of a low-frequency electromagnetic quantum and the amplitude of
   elastic scattering [compare Eq.(\ref{pole})].  Relations of this
   type provide a basis for the known infrared problem in QED, see
   e.g.\ \cite{LL4}.  Fig.~\ref{one} can be considered as a
   diagrammatic
   representation of Eq.(\ref{qu}).  The singular energy denominator
   $1/\omega$, which appears in Eq.(\ref{qu}), arises only in the
   external legs of the two Feynman diagrams shown in this picture.
   All energy denominators of all other diagrams include virtual
   energies of the atomic excitations, which are sufficiently high
   compared with the energy of the electromagnetic quantum.
   Correspondingly, all other diagrams, which are not shown in Fig.~\ref{one},
   are all finite in the limit $\omega \rightarrow 0$. This
   fact distinguishes the two diagrams in Fig.~\ref{one}, and
   guarantees that Eq.(\ref{qu}) is accurate for low frequencies.

   From Eq.(\ref{me}) one finds
     \begin{eqnarray}
       \label{squ}
       |f_\mathrm{abs}|^2
       =\left(\frac{2\pi \hbar}{m\omega}\right)^2
       \,(\bm{\epsilon} \cdot \bm{q})^2|f|^2
     \end{eqnarray}
     Averaging over possible orientations of the polarization vector
     one writes
     \begin{eqnarray}
       \label{ave}
       \left< \, (\bm{\epsilon} \cdot \bm{q})^2
       \right> =\frac{1}{3}\,q^2=\frac{2}{3}(1-\cos \theta)\,p^2~,
     \end{eqnarray}
     where brackets $\left<~\right>$ refer to the averaging procedure and
     $\theta$ is the scattering angle.  Eqs.(\ref{squ},\ref{ave})
     give
     \begin{eqnarray}
       \label{int}
       \int  \! \left< \,|f_\mathrm{abs}|^2 \right> d \Omega
       =\frac{2}{3}\left(\frac{2\pi \hbar \,p}{m\omega}\right)^2
     \! \!\int (1-\cos \theta )|f|^2\,d\Omega.~~
     \end{eqnarray}
     where the integration runs over the angles $\Omega$ of the
     scattered electron. The factor
     \begin{eqnarray}
       \label{tran}
       \sigma_\mathrm{tr}=\int (1-\cos \theta )|f|^2\,d\Omega~.
     \end{eqnarray}
     represents the transport cross section on the potential $U$.
     Eqs.(\ref{int},\ref{tran}) give
     \begin{eqnarray}
       \label{tra}
       \int  \! \left< \,|f_\mathrm{abs}|^2 \right> d \Omega
       =\frac{2}{3}\left(\frac{2\pi \hbar \,v}{\omega}\right)^2
       \sigma_\mathrm{tr}~.
     \end{eqnarray}
     Here $v=p/m$ is the velocity of the electron. The quantity on the
     left-hand side of Eq.(\ref{tra}) describes the probability of
     absorption of low-frequency quanta. The transport cross section
     on the right-hand side is related to elastic scattering. A close
     connection between low-frequency electromagnetic processes and
     elastic scattering is well known, see e.g.\  Ref.~\cite{LL4}.

     Deriving Eq.(\ref{tra}), we assumed that the potential $U$
     responsible for the electron scattering is localized within
     some finite volume.  Precisely this property allows one to
     distinguish the two Feynman diagrams in Fig.~\ref{one}. Otherwise,
     if the potential is spread all over an infinite volume, a mere
     concept of an external leg of the diagram would make no sense.

     We can specify the potential $U$ assuming that it is created by a
     single atom.  In that case Eq.(\ref{tra}) describes those events
     that take place during electron scattering by a single atom,
     being thus closely related to the single-atom approximation.

     \subsection{Kubo-Greenwood formalism}
     \label{Kubo-one-atom}
     Consider the conductivity of plasma, which is due to scattering
     of conducting electrons by atoms.  Within the Kubo-Greenwood
     formalism it can be written as
     \begin{eqnarray}
       \label{kubo}
       \sigma(\omega)= \frac{2 \pi n_\mathrm{a}e^2}{\omega}\int
       \left|  \langle \psi_{\bm{p}'} | \bm{\varepsilon}
       \cdot \bm{v }|\,\psi_{\bm{p}} \,\rangle \right|^2
   \left(  \mathsf{f}_{\bm{p}}-\mathsf{f}_{\bm{p}'}\right)
   \\ \nonumber
   \times \delta \left(\varepsilon_{\bm{p}}
     -\varepsilon_{\bm{p}'}-\omega\right)
   \frac{d^3p}{(2\pi \hbar)^3}
   \frac{d^3p'}{(2\pi \hbar)^3}~,
   \end{eqnarray}
   where $n_\mathrm{a}$ is the density of atoms, and
   $\mathsf{f}_{\bm{p}}$ is the Fermi distribution function
   for conducting electrons (which will be denoted by $\mathsf{f}$
   below)
   \begin{eqnarray}
     \label{distr}
     \mathsf{f}_{\bm{p}}=\frac{1}
     {\exp[ (\varepsilon_{\bm{p} }-\mu)/kT]+1}~.
   \end{eqnarray}
   The chemical potential $\mu$ here is related to the concentration
   of conducting electrons $n_\mathrm{c}$
   \begin{eqnarray}
     \label{n_c}
     2 \int \mathsf{f}\,\frac{d^3p}{(2\pi\hbar)^3}=n_\mathrm{c}~,
   \end{eqnarray}
   where the coefficient 2 accounts for two projections of spin.
   It should be noted that we have omitted contributions to Eq.(\ref{kubo})
   arising from atomic bound states. These contributions, which lead to bound-bound
   resonances and singularities near photoionization thresholds, are insignificant in
   the low-frequency region of concern herein.

   The first factor in the integrand in Eq.(\ref{kubo}) can be
   conveniently rewritten with the help of Eqs.(\ref{me},\ref{tra});
   the difference of the distribution functions in the integrand can
   be simplified using the low frequency approximation.  These
   transformations allow one to simplify the expression for
   conductivity Eq.(\ref{kubo}), reducing it to
   \begin{eqnarray}
     \label{simpli}
     \sigma(\omega)=\frac{2}{3}\,
       \frac{ n_\mathrm{a}e^2 }{ \omega^2 }\int v^3\sigma_\mathrm{tr}
       \left(-\frac{\partial
     \mathsf{f}}{\partial \varepsilon}\right)\frac{d^3p}{(2\pi\hbar)^3}
   \end{eqnarray}
   It is convenient to introduce the relaxation time
   $\tau_{\bm{p}}$ for conducting electrons, which is due to
   collisions with atoms
   \begin{eqnarray}
     \label{rela}
     \tau_{\bm{p}}=\frac{1}{v \,n_\mathrm{a}\sigma_\mathrm{tr} }~.
   \end{eqnarray}
   Clearly, it depends on the electron momentum via the velocity and
   the transport cross section. Eq.(\ref{simpli}) can be written in
   this notation in a transparent compact form
   \begin{eqnarray}
     \label{pres}
     \sigma(\omega)=\frac{ 2 e^2 }{3}\int
     \frac{v^2}{\omega^2\tau_{\bm{p}} }
       \left(-\frac{\partial
     \mathsf{f} }{\partial \varepsilon}\right)\frac{d^3p}{(2\pi\hbar)^3}~.
   \end{eqnarray}
   This is, in fact, the low-frequency limit of the Kubo-Greenwood
   formula in the single-atom approximation.

   There are two conditions that restrict a region of frequencies
   in which Eq.(\ref{pres}) is valid. Firstly, as was mentioned, the
   frequency must be sufficiently low. More precisely, this condition
   implies that the relevant scattering phases
   $\delta_l(\varepsilon)$, where $l$ is a typical orbital momentum,
   should not reveal significant variation in the interval of
   frequencies $\omega$
   \begin{eqnarray}
     \label{delta}
     \hbar \omega \left| \,
       \frac{\delta_l(\varepsilon)}{d\varepsilon}
         \, \right|\ll 1~.
   \end{eqnarray}
   Here $\varepsilon$ is a typical energy of conducting electrons.
   Secondly, the frequency is restricted from below by
   \begin{eqnarray}
     \label{below}
     \omega \tau_{\bm{p}} >1~,
   \end{eqnarray}
    $\bm{p}$ being a typical momentum of those
   electrons that give significant contributions to the
   conductivity in Eq.(\ref{pres}).  We will discuss this condition in
   detail after Eq.(\ref{result}). Here, let us mention
   briefly that the necessary high frequency specified by
   Eq.(\ref{below}) makes it certain that scattering processes on
   different atoms take place incoherently, as independent
   events; in other words, that the single-atom approximation is valid.
   \begin{figure}
     \centering \includegraphics[ height=6.7cm, keepaspectratio =
     true, angle =0]{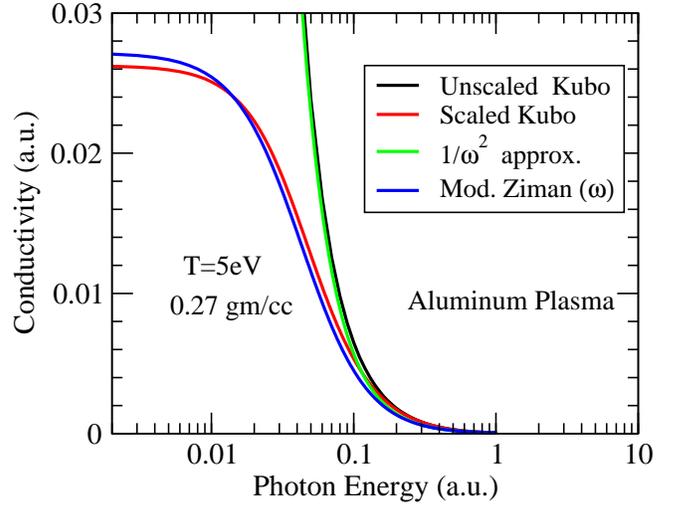}
   \caption{
     \label{two}
     The conductivity of an aluminum plasma at $T=5$ eV. Black line
     - the Kubo-Greenwood formula Eq.(\ref{kubo}). Green line -
     simplified Kubo-Greenwood formula Eq.(\ref{pres}), which predicts
     a second order pole $\sigma(\omega) \propto 1/\omega^2$. Red
     line - results from Ref.~\cite{johnson_guet_bertsch_04}, which
     used an interpolating procedure to extend the results of the
     Kubo-Greenwood approach to the static approximation described by
     the Ziman formula Eq.(\ref{ziman}). Blue line - prediction of
     Eq.(\ref{result}).}

   \end{figure}
   \noindent
   Eq.(\ref{pres}) predicts a simple $\propto 1/\omega^2$ behavior of
   the conductivity on frequency. If one ignores the restriction given
   in Eq.(\ref{below}) by taking the static limit in Eq.(\ref{pres})
     naively, then this equation clearly indicates that the
   conductivity has a second-order pole at $\omega=0$, as seen
   in Eq.(\ref{sec}).  Fig.~\ref{two}
   illustrates this statement by comparing calculations based on
   the complete Kubo-Greenwood formula Eq.(\ref{kubo}) with
   predictions of Eq.(\ref{pres}).  As an example, an
   aluminum plasma at temperature of 5 eV was taken in
   Fig.~\ref{two}.  The agreement between the two sets of calculations,
   shown in the black and green lines in Fig.~\ref{two},
     supports the validity of the approximations, which led to
     Eq.(\ref{pres}). The numerical code for calculations
     reported in the present work are based on the average-atom model and
     the numerical methods suggested in Ref.~\cite{johnson_guet_bertsch_04}.

   \section{Conductivity at ultra-low frequencies}

   \subsection{Ziman formula}
   \label{Ziman formula}

   Consider the static limit $\omega=0$.  The Ziman formula, which
   describes the conductivity due to electron-atom scattering reads
   \begin{eqnarray}
     \label{ziman}
     \sigma(0)=\frac{ 2e^2 }{3}\, \int v^2
     \tau_{\bm{p}}
       \left(-\frac{\partial
     \mathsf{f} }{\partial \varepsilon}\right)\frac{d^3p}{(2\pi\hbar)^3}~.
   \end{eqnarray}
   One observes its drastic distinctions from the result of the
   Kubo-Greenwood - type approach. Firstly, the Ziman formula
   Eq.(\ref{ziman}) gives a constant static limit for the
   conductivity, while the Kubo-Greenwood formula Eq.(\ref{pres})
   diverges at $\omega= 0$.  Secondly, in these two formulas the
   conductivity shows an opposite dependence on the relaxation time;
   Eq.(\ref{pres}) reveals an inverse dependence, $ \propto 1/
   \tau_{\bm{p}}$, while the Ziman formula (\ref{ziman}) predicts
   a linear dependence, $\propto \tau_{\bm{p}}$.

   To find an origin for these distinctions, let us note that deriving
   Eq.(\ref{pres}) we assumed that the electron momentum
   $\bm{p}$ is a good quantum number, which is changed only
   due to scattering on one given atom. Generally speaking, this
   assumption is incorrect. The momentum can be changed due to
   scattering on other atoms as well.

   To see the implications of this fact more clearly, let us note that the
   second-order pole $\propto 1/\omega^2$ in the conductivity arises
   as a direct consequence of the first-order pole $\propto 1/\omega$
   in the amplitude. The latter can be written as an integral
   \begin{eqnarray}
     \label{time}
     \frac{1}{\omega}=-i\int_{- {\infty} }^0\exp (-i\omega t) \,dt~.
   \end{eqnarray}
   where $|t|$ gives a period of time, which precedes an electron
   collision with the given atom. Eq.(\ref{time}) shows that deriving
   Eq.(\ref{pres}) one presumes that during all this period of time,
   which can be very large, up to infinity, the electron momentum
   remains constant.

   As a matter of fact, this is not true. The momentum can remain
   constant only over a finite period of time, which equals a typical
   interval of time between two subsequent collisions. This interval
   is measured by the relaxation time. This means that the relaxation
   time should necessarily produce the cut-off for the integral over
   time in Eq.(\ref{time})
   \begin{eqnarray}
     \label{cut}
     -i\int_{- {\infty} }^0\exp (-i\omega t) \,dt\rightarrow
     -i\int_{-\tau_{\bm{p} }} ^0 \exp (-i\omega t) ~.
   \end{eqnarray}
   The cut-off procedure can be fulfilled slightly differently and
   more conveniently, by introducing the cut-off function in the
   integrand
   \begin{eqnarray}
     \nonumber
     -i\int_{- {\infty} }^0 \exp (-i\omega t) \,dt\rightarrow
     -i\int_{-{\infty } }^0 \exp (-i\omega t-|t|/\tau_{\bm{p}})
    \\      \label{cuts}
    =\frac{1}{\omega+i/\tau_{\bm{p}} }\quad .
   \end{eqnarray}
   The pole at $\omega=0$, which exists on the left-hand side here, is
   replaced by a finite behavior of the right-hand side.

   These simple arguments show that the pole $\propto 1/\omega$ in the
   scattering amplitude and, correspondingly the second-order pole
   $\propto 1/\omega^2$ in the conductivity, are closely related to
   the single-atom approximation. The many-atomic events lead to the
   relaxation time, which erases this pole behavior. This argument is
   developed below, in Section \ref{resonance}.

   \section{Resonant states of conducting electrons}
   \label{resonance}

   According to Sec.~\ref{Ziman formula}, multiple scattering
   events should play an important role in the ultra-low frequency
   region.  In order to account for this scattering, let us start from
   a simple physical picture. If the electron has the momentum
   $\bm{p}$, then it keeps this momentum only for some finite
   period of time (relaxation time) because collisions with atoms in a
   plasma inevitably change it.  In the classical approximation this
   implies that only some finite part of the classical trajectory of the
   electron can be described by the initial momentum, while longer
   parts of the trajectory "forget" this momentum. Similarly, in
   quantum description the stationary quantum states, which describe
   the electron propagation in a plasma, cannot be characterized by the
   momentum.

   However, if the plasma is sufficiently transparent, i.e.\ the
   relaxation time is sufficiently large, then during long intervals
   of time the electron momentum on classical trajectories remains
   constant.  Consequently, the quantum states, which describe the
   electron propagation in a region outside atomic cores during
   moments of time separating consequent collisions, should look
   similar to conventional plane waves. The fact that collisions,
   which destroy the electron momentum, are essential can be accounted
   for by stating that a quantum state with the given momentum
   $\bm{p}$ exists only during a finite period of time.  In
   other words, the electron wave function of the conducting electron
   outside the atomic core of some atom resembles a conventional plane
   wave, but with the restriction that it exists only during a finite
   period of time that equals the relaxation time. Presuming that the
   relaxation time is large, one can say that this wave function is a
   quasistationary state, which is similar to a plane wave, but
   possesses a finite width $\Gamma_{\bm{p}}$ defined by the
   relaxation time
   \begin{eqnarray}
     \label{Gamma}
     \frac{  \Gamma_{ \bm{p} } }{2}=\frac{ \hbar }
     {\tau_{ \bm{p}} }~.
   \end{eqnarray}
   This identity, combined with Eq.(\ref{rela}), states simply that $
   \Gamma_{ \bm{p} }=2\hbar v n_\mathrm{a}\sigma_\mathrm{tr}$,
   which makes sense.

   The arguments just presented show that the electron wave function
   outside the atomic core of some atom can be written in a form
   \begin{eqnarray}
     \label{pw}
     |\bm{p},t\rangle =\exp\left( i(\bm{p} \cdot \bm{r} -
     \varepsilon_{ \bm{p} }t)-\frac{\Gamma_{ \bm{p} }}{2}t\right) ~.
   \end{eqnarray}
   This simple wave function possesses important physical properties.
   Firstly, it is close to a plane wave.  Secondly, its finite width
   accounts for multiple-scattering events; i.e. collisions with different atoms.
   The width of this resonant state is described by the relaxation
   time Eq.(\ref{Gamma}); the larger is the relaxation time, the
   closer is the wave function to a plane wave; exactly what one
   should expect when collisions are rare.


   The above argument can be developed further. If one wishes to consider the
   electron wave function in a close vicinity of a given atom, then
   the plane wave should be replaced by the wave function which takes
   into account the influence of the atomic potential $U(\bm
   {r})$. In other words, one needs to make in Eq.(\ref{pw}) a
   substitution $\exp( i \bm{p} \cdot \bm{r})\rightarrow
   \psi_{\bm{p} }(\bm {r})$. As a result, the wave
   function of a conducting electron, which takes into account the
   potential of a given atom, as well as scattering by other atomic
   particles, has the following form
   \begin{eqnarray}
     \label{Psi}
     \Psi_{\bm{p}}(\bm{r},t)=
     \psi_{\bm{p}}(\bm{r})\exp\left( i
     \varepsilon_{ \bm{p} }t - \frac{\Gamma_{ \bm{p} }}{2}t\right) ~.
   \end{eqnarray}
   Let us repeat, $\psi_{\bm{p}}(\bm{r})$ here is the
   wave function, which describes the electron behavior in the
   potential created by a single atom, while $\Gamma_{ \bm{p}
   }$ is the width, which describes the momentum relaxation due to
   scattering processes on all atoms.

   It is instructive to compare Eq.(\ref{Psi}) with a simple,
   classical idea of relaxation of the momentum. Consider for this
   purpose a value of the momentum averaged over the wave function
   Eq.(\ref{Psi})
   \begin{eqnarray}
     \label{av}
     \bm{P}(t)= \frac{1}{V}
      \int_{V} \Psi^*(\bm{r},t)\,
     \bm{p} \,\, \Psi(\bm{r},t)\,d^3r~.
   \end{eqnarray}
   Here $\bm{p}$ is the operator of the electron momentum and
   $V$ is a large, but finite volume, which makes a ratio in
   Eq.(\ref{av}) well defined, $V$-independent. From
   Eqs.(\ref{Psi}),(\ref{av}) one immediately finds that
   $\bm{P}(t)=\exp(-\Gamma_{\bm{p}}t/\hbar)\,
   \bm{P}(0)$. The above can be written in a more routine form
   \begin{eqnarray}
     \label{routine}
     \frac {d \bm{P}(t)}{dt}=-\frac{1}{\tau_{\bm{p}} }\,
   \bm{P}(t)~.
   \end{eqnarray}
   Clearly, this expresses a relaxation of the electron momentum in
   conventional classical terms; $\tau_{\bm{p}}$ plays here a
   role of the classical relaxation time, as one should have expected.
   Thus, a quantum description of the relaxation of the electron
   momentum based on the wave function Eq.(\ref{Psi}) reproduces a
   well-known conventional classical physical picture.

   Using the wave function Eq.(\ref{Psi}) in the Kubo-Greenwood
   formalism, one can follow a path outlined in Section
   \ref{Kubo-one-atom}. However, a well-known short-cut makes these
   calculations redundant.  A quasistationary nature of the wave
   function Eq.(\ref{Psi}) indicates that an amplitude of any resonant
   process involving this state acquires a conventional resonant
   energy denominator
   \begin{eqnarray}
     \label{res}
     \frac{1}{\Delta E+i\Gamma_{\bm{p}}/2}~.
   \end{eqnarray}
   Here $\Delta E$ is a deviation of energy from its resonant value,
   which is presumed to be low. In our case this deviation is defined
   by the frequency of the electromagnetic field $\Delta E=\hbar
   \omega$.  The resonant factor Eq.(\ref{res}) coincides with the one
   found in Eq.(\ref{cuts}), which underlines again a main physical
   idea; the multiple-scattering events allow the electron momentum to exist
   only during a finite period of time.

   The resonant amplitude Eq.(\ref{res}) always brings into the
   probability the resonant factor
   \begin{eqnarray}
     \label{prob}
     \frac{1}{\Delta E^2+\Gamma_{\bm{p}}^2/4}~,
   \end{eqnarray}
   which is often called the Breit-Wigner factor (in atomic physics
   this
   describes Lorentzian lines).

   Applying Eq.(\ref{prob}) to the process at hand one takes
   Eq.(\ref{pres}) and, making the substitution $1/\omega^2\rightarrow
   1/(\omega^2 +\Gamma_{\bm{p}}^2/4)$, arrives to the
   following result
   \begin{eqnarray}
     \label{result}
          \sigma(\omega)=\frac{ 2 e^2 }{3}\int
     \frac{ v^2  \tau_{ \bm{p}} } {  \omega^2\tau_{\bm{p}}^2+1 }
       \left(-\frac{\partial
     \mathsf{f} }{\partial \varepsilon}\right)\frac{d^3p}{(2\pi\hbar)^3}~.
   \end{eqnarray}
   As mentioned earlier, the above result can be obtained directly
   using the wave
   function Eq.(\ref{Psi}) in the Kubo-Greenwood formalism; however, the
   abbreviated derivation based on Eq.(\ref{prob}) makes the discussion more
   transparent.

   Eq.(\ref{result}) differs from Eq.(\ref{pres}), which was derived
   in the single-atom approximation, by the only physical fact; it
   accounts for  many-atom collisions. This distinction becomes
   crucial in the static limit, allowing Eq.(\ref{result}) to
   reproduce correctly the Ziman formula Eq.(\ref{ziman}). Thus, we
   return to the statement, which was mentioned several times
   previously; the many-atom events are very important for low
   frequencies. In contrast, in the high-frequency region
   $\omega\tau_{\bm{p}}>1$, Eq.(\ref{result}) reproduces
   Eq.(\ref{pres}), which is based on the simple single-atom
   approximation.  Thus, for $\omega\tau_{\bm{p}}>1$ the
   many-atom events become irrelevant, in agreement with discussion
   in Secs.~\ref{intro} and \ref{Kubo-one-atom}.

   Optical properties of a plasma are conveniently  described with the
   help of a complex conductivity, which allows one to define the
   complex refraction index. Using Eq.(\ref{result}) for the real part
   of the conductivity and applying the conventional Kramers-Kronig
   dispersion relation
   Refs.~\cite{kronig_1926,kramers_27,kronig_kramers_27,LL5,LL8}, one finds
   that its real and imaginary parts may be written side by side
   \begin{eqnarray}
     \label{re-im}
\left\{\! \!\begin{array}{l} \mathrm{Re} \\ \mathrm{Im} \end{array} \! \!\right\}
\sigma (\omega)
              =
              \frac{ 2e^2}{3} \!\! \int \!
\left\{\! \! \begin{array}{l} \mathrm{Re} \\ \mathrm{Im} \end{array}\!\! \right\}
     \frac{ i \,v^2 \tau_{\bm{p}}  } {  \omega\tau_{\bm{p}}  +i }
\left(\!-\frac{\partial
     \mathsf{f} }{\partial \varepsilon}\right) \frac{d^3p}{(2\pi\hbar)^3}.~~
   \end{eqnarray}
   Eq.(\ref{result}) is one of the main results of this work.  Its
   simple nature inspires a feeling that it could have been written
   without any discussion, as a simple convenient interpolation
   between the Ziman formula Eq.(\ref{ziman}) and the results of the
   Kubo-Greenwood approach Eq.(\ref{pres}). However, it is rewarding
   to realize that this result follows from a clear physical idea,
   which states that a conducting electron can possess a constant
   momentum only over a finite period of time.

   \section{Sum rule}
   \label{sum}

   Using Eq.(\ref{result}), one can calculate a simple but important
   integral
   \begin{eqnarray}
     \label{integral}
     \int_0^\infty \sigma(\omega)d\omega=\frac{ \pi}{3}\,\,e^2 \int v^2
    \left(-\frac{\partial
     \mathsf{f} }{\partial \varepsilon}\right)
    \frac{d^3p}{(2\pi\hbar)^3}~.
   \end{eqnarray}
   Rewriting here $d^3p= m^2 v \,d \, \varepsilon_{\bm{p}}\,
   d\Omega$ and integrating over the energy by parts one finds
   \begin{eqnarray}
     \label{parts}
     && \int_0^\infty \sigma (\omega) =
     \frac{\pi}{3}\,e^2m^2 \int v^3
    \left(-\frac{\partial
     \mathsf{f} }{\partial \varepsilon}\right) \frac{
    d\varepsilon\,d\Omega}
             {(2\pi\hbar)^3}
    \\ \nonumber
     && =\frac{\pi}{3} \, \frac{e^2}{ m }
     \int p^3
    \left(-\frac{ \partial \mathsf{f} }{ \partial \varepsilon} \right)
    \frac{ d\varepsilon\,d\Omega } {(2\pi\hbar)^3}=
     \pi\,e^2 \int p \,\,\mathsf{f} \,
    \frac{d\varepsilon\,d\Omega}
             {(2\pi\hbar)^3}
    \\ \nonumber
    && =\frac{\pi\, e^2}{m}\int \mathsf{f} \frac{d^3p}{(2\pi\hbar)^3}~.
   \end{eqnarray}
   Taking into account Eq.(\ref{n_c}), one finds that Eq.(\ref{parts})
   represents the known, conventional conductivity sum-rule
   \begin{eqnarray}
     \label{sum_rule}
     \frac{2}{\pi} \int_0^\infty\sigma (\omega) \, d\omega=\frac{n_\mathrm{c}e^2}{m}~.
   \end{eqnarray}
   The integration in Eq.(\ref{sum_rule}) includes the region of very
   high frequencies, above the limit in Eq.(\ref{freq}), which cannot
   be reliably covered by Eq.(\ref{result}). However, this region
   gives only a small contribution to the sum-rule because at high
   frequencies the integration in Eq.(\ref{sum_rule}) converges very
   well, as $\int d\omega/\omega^2$.
   The sum-rule Eq.(\ref{sum_rule}) supports validity of
   Eq.(\ref{result}).

   \section{summary}

   Let us summarize the main physical ideas. In a vicinity of a given
   atom the wave function of a conducting electron is
   strongly influenced by the potential of this atom. This fact makes
   it natural to presume that the problem can be formulated with the
   help of an average-atom model, which accounts for this variation and
   describes correctly the electron scattering on this atom.  However,
   the conductivity is related to processes of absorption and
   emission of electric field quanta during
   scattering of the conducting electron by a given atom.
   When the frequency of the field is ultra-low, the absorption and
   emission take place in a region located far away from the atom.
   The electron wave function in this region is necessarily influenced
   by potentials of other atoms.  As a result, the lower
   the frequency is, the more important are the many-atom events.
   Thus, the {\it naive} single-atom approximation inevitably breaks down
   in the static limit, where multiple scattering become crucial.

   From the first glance the necessity to account for multiple scattering
   should make things much more complicated for the theory. There is
   though an important simplification. The many-atom events manifest
   themselves mainly via a restriction, which they put on the period
   of time during which the conducting electron can possess a given
   momentum.  Henceforth, one can account for these events by stating
   that the wave function of a conducting electron is a
   quasistationary state, which exists only during a large, but finite
   period of time, which equals the relaxation time for the momentum.
   This idea can be expressed in terms of the quasistationary state
   Eq.(\ref{Psi}), which describes the conducting electron.  As a
   result, it becomes possible to account for multiple scattering
   within the formalism of the average-atom approximation, which greatly
   simplifies the problem.

   Applying this idea within the framework of the Kubo-Greenwood
   formalism, we find that the conductivity is given by
   Eq.(\ref{result}), which possesses several interesting properties.
   \begin{enumerate}
   \item In the static limit it reproduces the Ziman formula
     Eq.(\ref{ziman}).
   \item In the high-frequency region it is reduced to the
     Kubo-Greenwood type formula Eq.(\ref{pres}) derived within the
     {\it naive} average-atom approximation.
   \item It satisfies the conventional sum-rule Eq.(\ref{sum_rule}).
   \item It is formulated in terms of physical quantities, which can
     be evaluated in the average-atom approximation that is convenient for
     applications.
   \end{enumerate}
   Our starting point was Eq.(\ref{qu}), which relates the elastic
   scattering amplitude with the amplitude of absorption (emission)
   of low frequency quanta.  The latter gives a particular example of
   a general property of QED, which allows one to express any
   radiation process with soft quanta via a purely elastic scattering
   process, see e.g.\  \cite{LL4}.  Starting from Eq.(\ref{qu}), we
   derive Eq.(\ref{pres}) using a single-atom approximation, then
   taking into account many-atom events upgrade it to
   Eq.(\ref{result}) which, in the static limit, reproduces the Ziman
   formula Eq.(\ref{ziman}) for Ohm's law. Thus, the well known,
   conventional Ohm's law may be considered as a direct consequence of
   general, fundamental infra-red properties of QED.

\begin{figure}
  \centering \includegraphics[ height=6.5cm, keepaspectratio = true,
  angle =0]{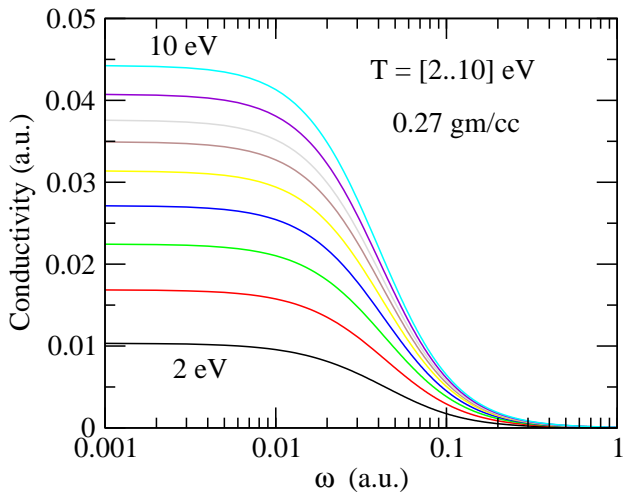}
\caption{ \label{three}
  The conductivity of an aluminum plasma at different temperatures.
  Calculations are based on Eq.(\ref{result}). The parameters used in this
  formula are evaluated with the help of the
  approach of Ref.~\cite{johnson_guet_bertsch_04}. }
\end{figure}
   \noindent

   Using Eq.(\ref{result}) and evaluation the necessary average-atom
   quantities $\tau_{\bm{p}}$ and
   $\mathsf{f}_{\bm{p}}$ with the help of the model of
   Ref.~\cite{johnson_guet_bertsch_04} we calculated the conductivity of
   the aluminum plasma at different temperatures.
   The results are
   shown in Figs.~\ref{two} and \ref{three}. In the one-atom
   approximation the conductivity is divergent as $\sigma\propto
   1/\omega^2$, as shown in the black and green lines in Fig.~\ref{two}.
   To avoid this unphysical divergence
   Ref.~\cite{johnson_guet_bertsch_04} suggested a particular
   interpolating
   procedure, shown by a blue line in Fig.~\ref{two}, which brings
   the conductivity to the Ziman formulae Eq.(\ref{ziman}) in the
   static limit.  Eq.(\ref{result}) provides more rigorous treatment
   of the conductivity at low frequencies, which does not rely on an
   interpolation.  It is satisfying that the two approaches give close
   numerical results; compare the red and blue lines in Fig.~\ref{two}.

   Fig.~\ref{three} presents results of a series of calculations based
   on Eq.(\ref{result}) for different temperatures of the plasma.  The
   increase of the conductivity with temperature reflects an increase of
   the concentration of conducting electrons. The rapid decrease of the
   conductivity in the high-frequency region underlines the important
   role played by low frequencies; as was mentioned, the low-frequency
   region gives a dominant contribution to the sum-rule Eq.(\ref{sum_rule}).

     In conclusion, it is shown that Eq.(\ref{result}) successfully
     describes plasma conductivity at low-frequencies.

  \acknowledgments

 One of us (MK) is thankful for the hospitality of the staff of the
 Physics Department of the University of Notre Dame, where this
 work was initiated and acknowledges financial support from the
 Australian Research council and the Australian Academy of Sciences.
 WJ is thankful for partial support from the School of Physics at the
 University of New South Wales and for partial support from NSF grant
 No.\ PHY-0456828. Helpful discussions with Claude Guet are also gratefully
 acknowledged.


\begin{thebibliography}{31}
\bibitem[{end()}]{endnote1}
\bibinfo{note}{The field can slowly vary in space, but a typical wave vector
  $k$, which describes this variation should be small, $k\ll \omega/v$, where
  $\omega$ is the frequency of oscillations of the electric field and $v$ is a
  typical velocity of conducting electrons.}
\end{thebibliography}

\end{document}